# Spectral Convolutional Neural Network Chip for In-sensor Edge Computing of Incoherent Natural Light

Kaiyu Cui[#*1], Shijie Rao[#1], Sheng Xu[1], Yidong Huang[*1], Xusheng Cai[2], Zhilei Huang[2], Yu Wang[2], Xue Feng[1], Fang Liu[1], Wei Zhang[1], Yali Li[1], and Shengjin Wang[1]

[1]Department of Electronic Engineering, Tsinghua University, Beijing, China.

[2]Beijing Seetrum Technology Co., Beijing, China.

\* Corresponding author. Email: <u>kaiyucui@tsinghua.edu.cn</u>, <u>yidonghuang@tsinghua.edu.cn</u>

[#] These authors contributed equally to this work

**Abstract:**

Convolutional neural networks (CNNs) are representative models of artificial neural networks (ANNs). However, the considerable power consumption and limited computing speed of electrical computing platforms restrict further CNN development on edge devices. Optical neural networks are considered next-generation physical implementations of ANNs, but their capabilities are limited by on-chip integration scale and requirement for coherent light sources. This study proposes a spectral convolutional neural network (SCNN) of incoherent natural light by an optical convolutional layer (OCL) and a reconfigurable electrical backend. The OCL is implemented by integrating very large-scale, pixel-aligned spectral filters on a CMOS image sensor on a 12-inch wafer, facilitating highly parallel spectral vector-inner products of incident light. It accepts broadband incoherent natural light containing two spatial and one spectral dimension directly as input with the function of matter meta-imaging. This unique optoelectronic framework empowers in-sensor optical analog computing at extremely high energy efficiency because the OCL is driven by the energy of the information carrier, i. e. natural light. To the best of our knowledge, this is the first integrated optical computing utilizing natural light. We employ the same SCNN chip for completely different real-world complex tasks, and achieve accuracies of over 96% for pathological diagnosis and almost 100% for face anti-spoofing at video rates. The SCNN framework has an unprecedented new function of substance identification, provides a feasible optoelectronic and integrated optical CNN implementation for edge devices or cellphones, providing them with practical and powerful edge computing abilities and facilitating diverse applications, such as intelligent robotics, industrial automation, medical diagnosis, and remote sensing.

**Introduction:**

Artificial neural networks (ANNs) have demonstrated powerful abilities across numerous applications, such as the burgeoning ChatGPT[1] and AIGC[2], and have altered many aspects of modern society. Because vision is the most important method for both humans and machines to perceive the world, among different ANNs, convolutional neural networks (CNNs) inspired by biological vision for image processing have become one of the most commonly used ANN architectures[3]. Owing to the convolutional layers that enable CNNs to extract high-level features from raw image data and significantly reduce parametric complexity[3,4], CNNs have achieved considerable success in image recognition[5], segmentation[6], and detection[7] tasks. However, the convolutional processing of the network dominates the processing time and computing power. This leads to significant computing cost challenges and severe limitations for CNNs on leading high-performance electronic computing platforms, such as graphic process units (GPU), as reflected by Moore's law[8]. The huge computational cost severely limits the deployment of CNNs on portable terminals for edge computing.

Optical neural networks (ONNs), or optical neuromorphic hardware accelerators, have been regarded as one of the most promising next-generation parallel-computing platforms to address the limitations of electronic computing, with the distinct advantages of fast computational speed, high parallelism, and low power consumption[9-14]. Existing works on ONNs have achieved fully connected neural networks (FCNs) based on the Reck design[15-19] or diffractive deep neural network ($D^2NN$)[20-25] and optical CNNs (OCNNs) or optical convolutional accelerators by further introducing wavelength division multiplexing[26-29], attaining extraordinary computing speed with low power consumption. However, existing on-chip OCNNs hardly accept broadband incoherent natural light. The requirement for a coherent light source limits the scale of optical matrix multiplication[30] and are insufficient for two-dimensional (2D) convolution calculations. Moreover, in these works[11,24,26,27,31-33], broadband incoherent natural light is usually captured by digital cameras and then encoded to coherent light for optical computing (Fig. 1a), which not only degrades the energy efficiency but also loses the light field features containing rich matter information, such as spectrum, polarization and incident angle. Especially, the spectral features that can identify the composition of matter for complex vision tasks cannot be directly introduced into OCNNs.

In this work, we propose and demonstrate a spectral convolutional neural network (SCNN) based on an optoelectronic computing framework that accepts broadband incoherent natural light directly as input (Fig. 1b). Hybrid optoelectronic computing hardware with an optical convolutional layer (OCL) and a reconfigurable electrical backend is employed to leverage optical superiority without sacrificing the flexibility of digital electronics[19,20,28-30,34-37]. The OCL works as the input and the first convolutional layer. Our proposed OCL utilizes very large-scale, pixel-aligned integration of spectral filters on a CMOS image sensor (CIS), as is shown in Fig. 1c and Fig. 1d. Here, the spectral filters can utilize dispersive nanostructures or material with spectral modulation abilities. In this work, we provide two implementations of the spectral filters. The first one is based on metasurfaces which provide better spectral modulation capabilities (Fig. 1c). The second one is achieved by pigments with mass production on a 12-inch wafer (Fig. 1d). The weights of the OCL are encoded on the transmission responses of the spectral filters. It should be noted that previous hyperspectral imaging works adopted spectral filters as the sensing matrix and got the compressively sensed hyperspectral images[38-41]. After capturing, the hyperspectral images require post-processing of spectral reconstruction and further spectral analysis. In these systems, the spectral filters are designed to achieve high spectral resolution and the post-processing of the captured data requires huge computational cost, which is incapable of applying on edge computing. In this work, the spectral filters are designed to be the first layer of the neural network. Their transmission responses work as weights of the layer rather than the sensing matrix. Therefore, we only need very few tailored spectral filters to achieve real-world applications at high efficiency because accurate spectral reconstruction is not required thus achieving edge computing. In this work, only 9 different spectral filters are designed for the SCNN. More detailed comparison is described in Supplementary Note 1.

After natural light transmits through the broadband spectral filters, CIS is used to detect the light intensity at different spatial locations (Fig. 1b), which sums the energy of the transmitted light along the wavelength axis (i.e., the spectrum) at each image pixel, similar to the functions of cone cells in the human eye. Therefore, the CIS and spectral filters form an analog OCL with high spatial resolution and process natural images directly without explicit image duplication. As the OCL facilitates a highly parallel vector inner-product that is driven by the energy of input natural light and completed during the light field sensing process, it achieves real-time in-sensor computing. In this work, the OCL achieves about 21.0 tera operations per second (TOPS) in computing speed and 96% reduction in data throughput, so that the computational load of the electrical backend can be significantly reduced. Besides, incoherent natural light includes two spatial dimensions and one spectral dimension, the composition of matter cand be identified and mapping of its distribution in space can be realized by the SCNN, which start a new paradigm for matter meta-imaging (MMI) beyond human eyes. To verify the capabilities of the proposed SCNN framework, we conducted several real-world complex vision tasks at video-rate with the same SCNN chip, including pathological diagnoses with over 96% accuracy and anti-spoofing face recognition with 100% accuracy. Our implementation enables low-cost mass production and integration in the edge devices or cellphones of the proposed SCNN. Therefore, the proposed SCNN provides new MMI vision hardware and edge computing abilities for terminal artificial intelligence systems.

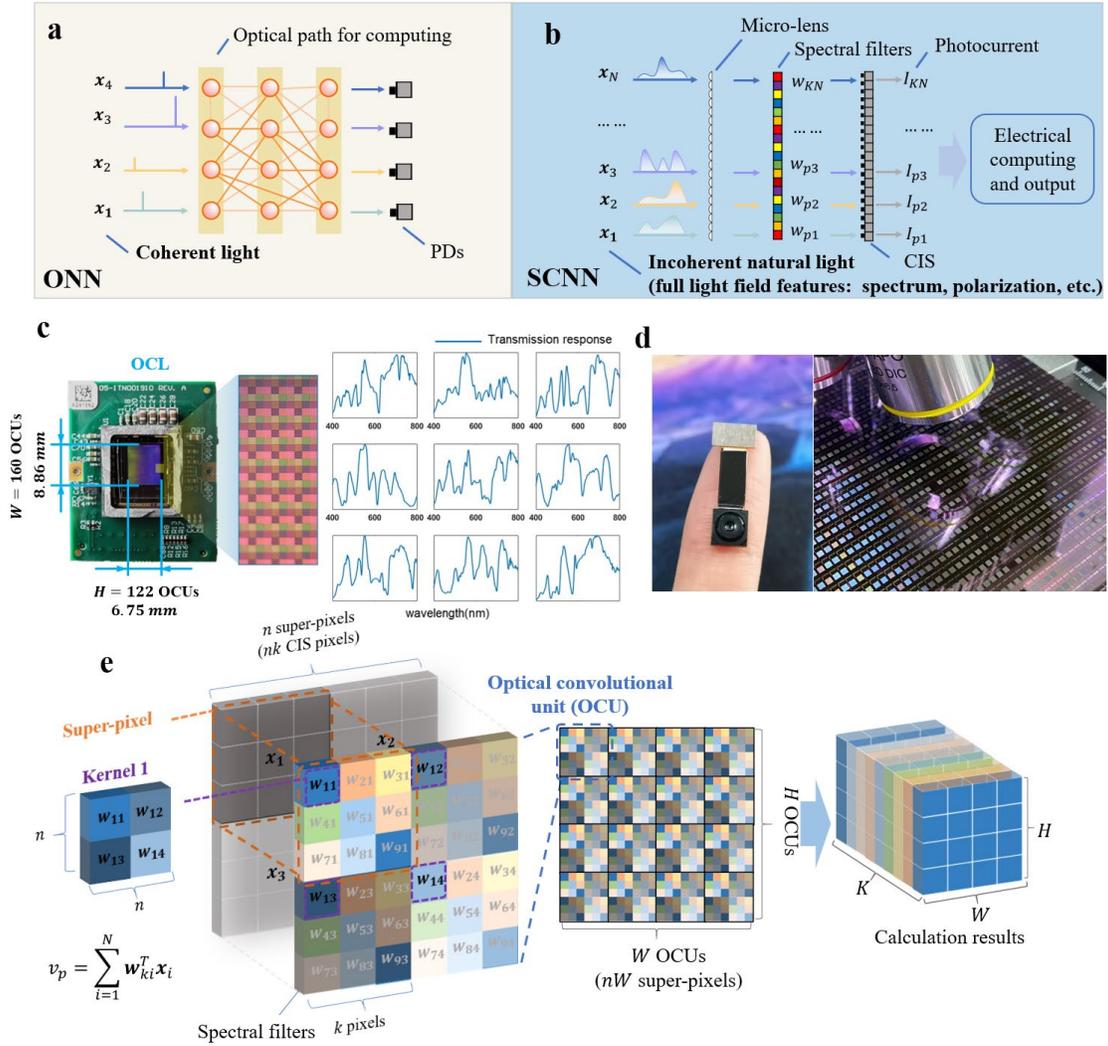

**Fig. 1 Principles of the proposed SCNN. a,** Existing ONNs are based on coherent light sources for computing. They are incapable of broadband light field sensing and in-sensor computing. **b,** In our design, we implemented an SCNN by integrating very large-scale spectral filters on CIS. Our SCNN can accept incoherent natural light and perform analog 2D convolution calculations directly. **c,** The metasurface-based OCL integrates pixel-aligned metasurface units on a CIS. **d,** The pigment-based OCL is fabricated by lithography on a 12-inch wafer. **e,** The working principles of our OCL. One OCL contains an $H \times W$ array of identical OCUs and each OCU has $K$ convolutional kernels, resulting in calculation results of size $H \times W \times K$.

# Results

**SCNN Architecture.** Our proposed SCNN consists of various spectral filters integrated on a CIS functioning as an on-chip analog OCL, followed by several electrical network layers (ENLs), as shown in Fig. 1b. Here, the spectral filters are designed to modulate light at different spectral and spatial points, which applies the convolutional kernel weights. Each spectral filter is completely aligned to a CIS pixel. $K = k \times k$ CIS pixels constitute a super-pixel and $N = n \times n$ super-pixels form an optical convolutional unit (OCU), as is shown in Fig. 1e. Furthermore, the entire OCL is an array of $H \times W$ OCUs. Because the OCUs are all identical, they perform spatial parallel analog 2D convolution calculations at different locations with megapixels.

Taking one OCU as an example (Fig. 1e), it has $K = k \times k$ convolutional kernels of size $n \times n$ and covers $n \times n$ super-pixels. The $p$-th ($1 \leq p \leq K$) kernel has $N = n \times n$ weight vectors $w_{p1}(\lambda), w_{p2}(\lambda), \ldots, w_{pN}(\lambda)$, where $w_{pi}(\lambda) = t_{pi}(\lambda) r(\lambda)$ is determined by the transmission response $t_{pi}(\lambda)$ of the $i$-th filter in the kernel and the quantum efficiency $r(\lambda)$ of the CIS. Assuming that the input visual information represented by the superpixel is $x_i(\lambda)$ ($1 \leq i \leq N$), the calculation result $v_p$ of the kernel is as follows:

$$v_p = \sum_{i=1}^{N} I_{pi} = \sum_{i=1}^{N} \int_{\lambda_1}^{\lambda_2} x_i(\lambda) w_{pi}(\lambda) d\lambda = \sum_{i=1}^{N} w_{ki}^T x_i \tag{1}$$

where $I_{pi}$ denotes the electrical signal output of the CIS pixel under the $i$-th filter in the kernel. Each OCU contains $K$ kernels, and the OCL is a grid of $H \times W$ identical OCUs. Assume that the $p$-th kernel in the OCU located at $h, w$ ($1 \leq h \leq H, 1 \leq w \leq W$) has the output $v_{(h,w)p}$. Then, the 2D convolutional results of the OCL are:

$$F = \{v_{(h,w)p}\} \in R^{H \times W \times K} \tag{2}$$

Therefore, OCL has $K$ convolutional kernels of size $n \times n$ and stride $n \times n$. The input visual signal has a spatial resolution of $nH \times nW$ and $C$ spectral channel, which is equivalent to having $nH \times nW \times C$ voxels. $C$ is determined by the sampling points in the spectral dimension. We assume that the light is locally homogeneous in one superpixel.

The output feature map of the OCL has $H \times W$ spatial points and $K$ channels. As usually $K \ll C$, the OCL can greatly compress the information in the spectral domain.

After in-sensor computing by the OCL, the output feature map is sent to the trained ENLs, which can comprise various ANN architectures such as FCNs and CNNs. Although the tailored OCL hardware is fixed after fabrication in our SCNN framework, its kernel size $n$ and number of kernels $K = k^2$ can be reconfigured as well as $k \cdot n$ is fixed to the size of the OCU. A larger $n$ leads to better capabilities of extracting spatial features and a larger $k$ means more powerful spectral sensing abilities. Therefore, there is a trade-off between spatial and spectral features. We can choose the optimal value for $k$ and $n$ based on the actual needs of a specific task. Moreover, the ENLs can be changed and trained dynamically to suit different objectives. For example, in our disease diagnosis and face anti-spoofing tasks, we employed two different ENLs sharing the same OCL to perform pixel- and image-level predictions. Therefore, our SCNN framework combines the advantages of OCL by providing ultrafast sensing and processing of spatial and spectral features of natural images and the flexibility of ENLs with reconfigurable network designs for different tasks, enabling real-time MMI for different machine intelligent systems. Particularly, the OCL significantly reduces the computational load and data throughput of the electrical backends. The whole system can run in real-time without the need for GPU. Therefore, the entire system is efficient and compact, which open the way for edge computing applications.

**Metasurface based SCNN Chip.** In this work, we provide two implementations of the spectral filters for the SCNN. The first one is based on metasurfaces which provide flexible designed spectral modulation for the kernel weights of the OCL. Since different functions and applications require distinct metasurface designs to achieve the best results, we propose a gradient-based metasurface topology optimization (GMTO) algorithm to achieve an application-oriented metasurface design for different tasks such as thyroid disease diagnosis and anti-spoofing face recognition (Fig. 2a). Here, we first adopted freeform-shaped meta-atom metasurfaces[41] to generate millions of different metasurface units and arranged all the metasurfaces into a 2D array. Thus, each metasurface unit can be uniquely represented by a pair of coordinates $(p, q)$. To design $N$ metasurfaces, the objective can be considered a function of $2N$ independent variables: $L(p_1, q_1, ..., p_N, q_N)$. We then utilized the GMTO algorithm to find the minimum points of $L(p_1, q_1, ..., p_N, q_N)$, obtaining the optimized design (see Supplementary Note 3 for details).

We found that OCL, designed by GMTO, could extract discriminating features with as few as nine kernels for live human skin and the thyroid tissue. The visualization results by principal component analysis (PCA)[42] are shown in Fig. 2b and Fig.3c, recpectively. Fewer kernels enable higher feature compression capability, higher spatial resolution, and lower computing costs for ENLs. Particularly, compared with our previous hyperspectral imaging works[35-38], SCNN use very small number of metasurface units and providing an ONN-based approach for hyperspectral sensing, effectively avoiding the need for as many metasurface units as possible for high precision spectral reconstruction (see Supplementary Note 1 for details). Finally, we implemented the OCUs with $H = 122$ and $W = 160$ by integrating millions of pixel-aligned metasurface units on top of a CIS (see Methods for details). The scanning electron microscopy (SEM) images of the fabricated metasurfaces are shown in Fig. 2a.

As is mentioned above, the size and number of convolutional kernels can be reconfigured. For example, the OCL shown in Fig. 1c can also be regarded as having 1 ($k = 1, K = k^2 = 1$) convolutional kernel of size $3 \times 3$ and stride $3 \times 3$ ($n = 3$). In this configuration, we need to sum the outputs of all of the CIS pixels in one OCU to generate an output feature map of size $160 \times 122 \times 1$. We find this configuration performs worst in experiments because spectral features are more important than spatial features in the two applications. Therefore, we adopt the configuration of 9 convolutional kernels of size $1 \times 1$ and stride $1 \times 1$ to conduct further experiments.

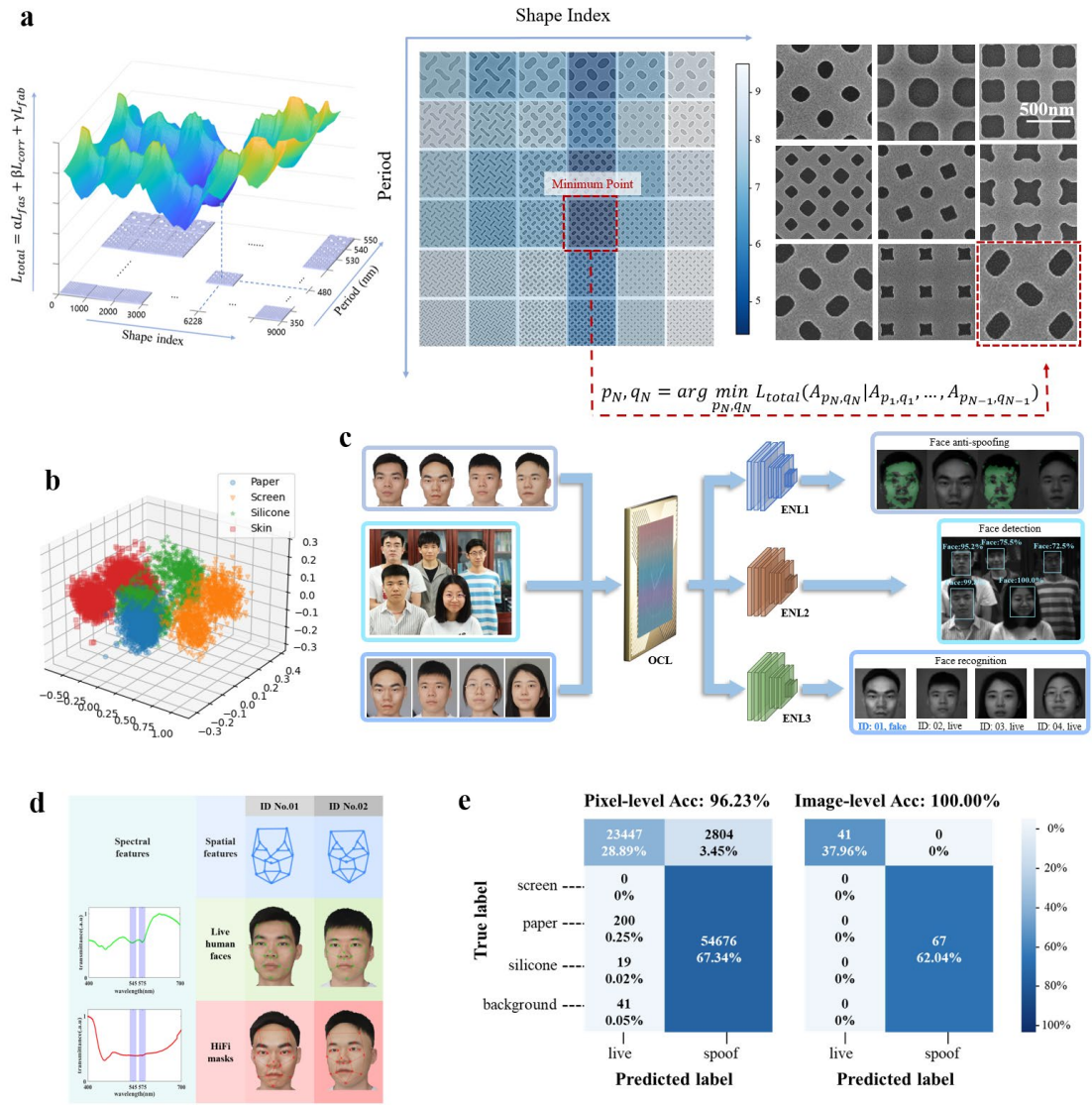

**Fig. 2 Metasurface based SCNN chip can be used for multiple vision tasks related to face recognition. a,** The GMTO algorithm is achieved by finding the minimum point of the designed loss function. **b,** Spectral feature extraction results of the OCL visualized by PCA. Live skin and three spoof materials are separated. **c,** The optical convolutional layer (OCL) has 9 kernels with size $1 \times 1$. By changing the electrical network layers (ENLs), the same SCNN chip can be trained to complete face anti-spoofing, face detection and face recognition tasks. **d,** Our SCNN chip can combine spectral features with spatial features and perform reliable anti-spoofing face recognition. **e,** Confusion matrix for the pixel-level and image-level liveness detection results.

To test the capabilities of the proposed SCNN framework, we employed the proposed SCNN for face anti-spoofing (FAS) to verify its performance. Nearly all of the current face recognition systems can be deceived by high-fidelity (HiFi) silicone masks, posing a great risk to privacy and security. However, when powered by our MMI, discriminative features can be extracted to detect HiFi masks. We captured images and obtained a test set containing 108 test samples from 31 different people, including several HiFi silicone masks, under natural light, and evaluated the performance of our SCNN chip. The results are shown in Fig. 2c and Fig. 2d. We can observe that our SCNN chip can effectively recognize live pixels, which are marked in green. Fig. 2e shows the confusion matrix of the SCNN for all the test samples. The SCNN framework achieved 100% and 96.23% accuracy in image- and pixel-level liveness detection on our test dataset, demonstrating that our SCNN chip can achieve high reliability in anti-spoofing liveness detection applications (more results can be found in Supplementary Note 4). These results indicate the considerable potential for FAS systems.

Furthermore, we employed the designed SCNN chip to perform real-time anti-spoofing pixel-level liveness detection at different video frames. In this experiment, the entire system was run on a traditional Intel Core i5-6300HQ CPU, and the frame rate of the results was only limited by the CIS exposure time. The HiFi silicone masks can be easily detected at pixel-level (more results can be found in the Supplementary Video 1). Thus, the proposed SCNN framework is expected to be widely used in the real-world applications of MMI. By simply redesigning and retraining the ENLs according to the needs of specific tasks, the function of the SCNN can be customized, such as face detection and recognition, as shown in Fig. 2c (more details of the redesigned ENLs can be found in Supplementary Note 5). The results show that the SCNN can accurately predict the locations of faces and achieve face recognition. This experiment indicates that the final output of the SCNN is highly customizable. The SCNN can flexibly adapt to various advanced CV tasks at video rates by simply changing and retraining the ENLs.

In addition to face anti-spoofing, we conducted automatic thyroid disease diagnosis experiments. The samples included normal thyroid tissue and tissues from four different diseases: simple goiter, toxic goiter, thyroid adenoma, and thyroid carcinoma. As shown in Fig. 3a, natural images of thyroid histological sections were first detected and processed using OCL. The feature maps output by the OCL are further processed by the ENLs to output the image-level thyroid disease classification results. Finally, the pixel-level disease detection results were output by other ENLs (see Supplementary Note 6 for details about the network).

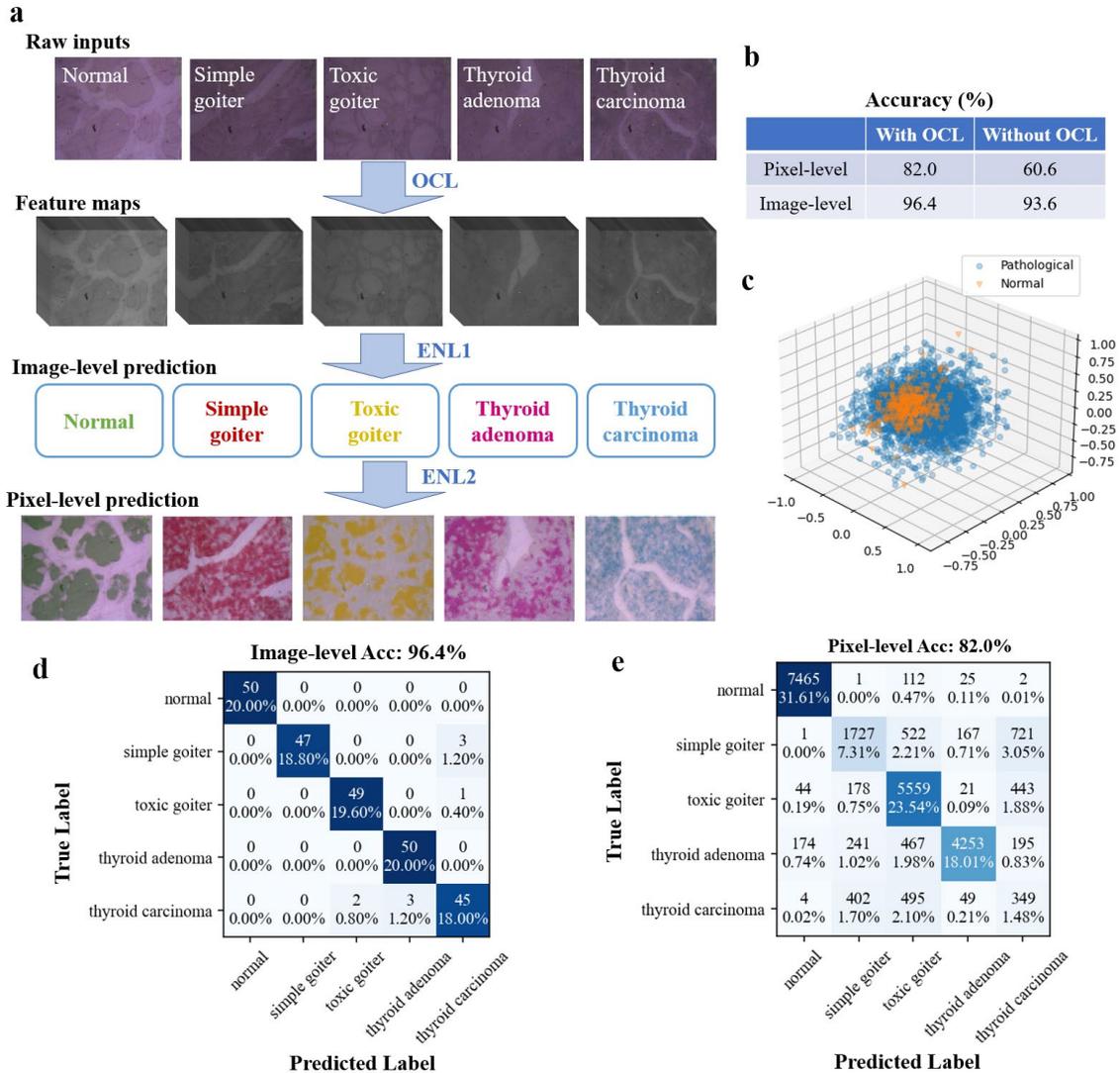

**Fig. 3. Experimental results of thyroid histological section diagnosis by the Metasurface based SCNN. a,** We exploit our SCNN to sense the raw datacube of thyroid histological section through a microscope. After the data are processed by the optical convolutional layer (OCL) and electrical network layers (ENLs), thyroid disease is automatically determined via image-level prediction. After the data are processed further by additional ENLs, the potential pathological areas are labeled in different colors via pixel-level prediction. **b,** Without OCL, the classification accuracy based on the same monochromatic sensor decreases considerably for both image- and pixel-level predictions. c, The spectral features from OCL can be visualized by PCA. Normal and pathological tissues are separated. **d,** Confusion matrix of the image-level thyroid pathology classification results of the SCNN chip on the test set. Our SCNN chip achieves 96.4% accuracy. **e,** Confusion matrix of the pixel-level results. Our SCNN chip achieves 82.0% accuracy.

Fig. 3d and Fig. 3e show that our SCNN framework can diagnose these four thyroid diseases, achieving an image-level testing accuracy of 96.4%. Moreover, the SCNN chip automatically labeled the potential pathological areas in different colors at high spatial resolution. To study the role of the OCL, we conducted another experiment by replacing the OCL with a CIS without metasurfaces. After repeating the same ENLs training procedure, the image-level prediction accuracy decreased from 96.4% to 93.6%, and the pixel-level prediction accuracy decreased from 82.0% to 60.6% (Fig. 3b). The performance is much worse than using the OCL because OCL provides extra spectral sensing capabilities. Therefore, for the vision tasks related to spectral information, we need hyperspectral images rather than RGB images or grayscale images to get a good performance. If we complete the whole process by capturing data using a hyperspectral camera and implementing all neural network layers on the electrical computing platform, then we can get similar results compared with SCNN. However, the hyperspectral cameras usually have a very high cost and need time to scan a hyperspectral image. Moreover, the storing and processing cost of a hyperspectral image on an electrical computing platform is also very high (see Supplementary Note 2 for details). Therefore, conventional hyperspectral camera is not practical to be used in real-time edge-computing applications, while the SCNN provides a simple but highly effective way to sense and process hyperspectral images for various portable terminals.

**Pigments based SCNN Chip with Mass Production.** Besides metasurface-based spectral filters, we have also achieved the mass production of the SCNN on a 12-inch wafer utilizing pigments as spectral filters. The spectral filters are achieved by mixing several pigments with different organic solvents including ethyl acetate, cyclohexanone, and propylene glycol methyl ether acetate (PGMEA). The 12-inch wafer of the fabricated chips taped by lithography is shown in Fig. 4a. Each chip is only about $3 \times 3.5\ mm^2$ and can be integrated into any mobile device such as a smartphone to enable MMI. Focused ion beam-scanning electron microscope (FIB-SEM) image of the SCNN chip is shown in Fig. 4d. Each pigment-based filter is precisely aligned to a CIS pixel.

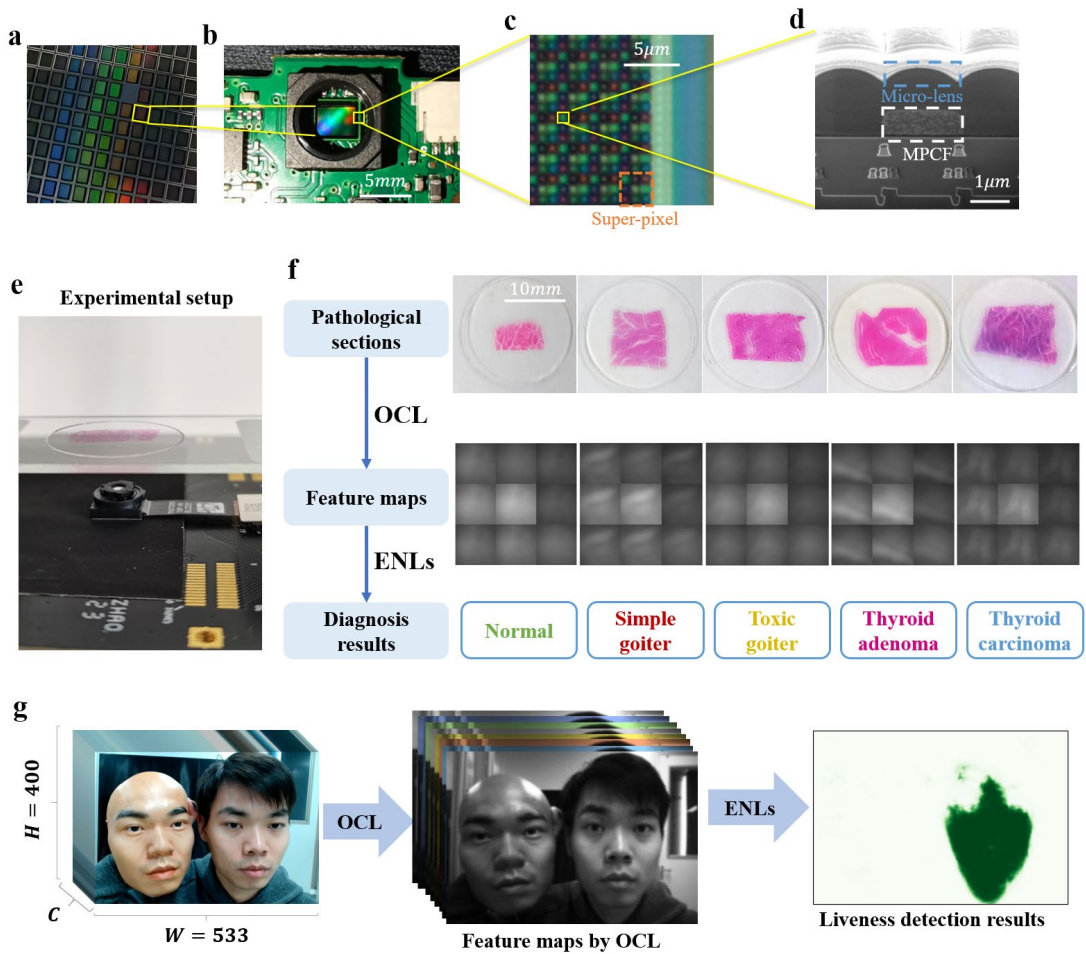

**Fig. 4. SCNN chip implemented by utilizing pigments as spectral filters and achieving mass production on a 12-inch wafer. a,** The fabricated SCNN chips on a 12-inch wafer by lithography. **b,** A tiny camera equipped with the SCNN chip. It can achieve in-sensor edge computing and spectral sensing. The size of the SCNN chip is only about

$3 \times 3.5\ mm^2$ and the size of the whole camera is about $6.5 \times 7\ mm^2$ **c,** A microscope image of the fabricated SCNN chip. It has 9 convolutional kernels of size $1 \times 1$ and stride $1 \times 1$. A super-pixel contains 9 CIS pixels. **d,** The focused ion beam-scanning electron microscope (FIB-SEM) image of the SCNN chip. One CIS pixel is covered by a pigment-based spectral filter and a micro-lens. The fabrication process is completely standard semiconductor lithography process. **e,** We place the thyroid pathological sections right above the lens without a microscope. **f,** The sections and the corresponding feature maps outputted by OCL. **g,** The face anti-spoofing results of the pigment-based SCNN.

We selected 9 different pigments to form the spectral filters from several candidates that are compatible with lithography to make the differences between different targets in the feature maps outputted by the OCL as large as possible. Lithography enables large-scale integration of spectral filters, and the SCNN chip has a total of $400 \times 533$ ($H = 400, W = 533$) superpixels. Therefore, the size of the feature map output by OCL is $400 \times 533 \times 9$. The spatial resolution is sufficient for most computer vision tasks, and OCL empowers massive parallel analog computing.

The fabricated chip is packaged into a tiny camera, as shown in Fig. 4b. The size of the camera is approximately $6.5 \times 7\ mm^2$. We placed the pathological thyroid sections immediately above the camera lens without any microscope, which is impossible for traditional pathological diagnosis. Natural images of thyroid histological sections were first obtained and processed using OCL. The feature maps output by the OCL are then further processed by the ENLs to output the image-level thyroid disease classification results. The camera can capture only a blurry image rather than a sharp image showing clear textures since a microscope is not used. Some samples of pathological sections and their feature maps outputted by the OCL are shown in Fig. 4f. The feature maps display few spatial features. However, we still reach a classification accuracy of 96.46%. Furthermore, we also conducted another experiment by replacing OCL with CIS without pigment-based filters to study the role of OCL. After repeating the same data collection and ENL training procedure, the classification accuracy decreased from 96.46% to 47.09%. The tiny size of the finished camera allows it to be integrated into various medical instruments such as laparoscopes. Thus, the proposed SCNN framework shows considerable potential as an ancillary diagnostic tool in clinical medicine and might assist doctors in precisely localizing lesions in real-time during surgery.

We have also achieved the face anti-spoofing task using the pigment-based SCNN (Fig. 4g). The confusion matrix of the classification results and more experimental results can be found in Supplementary Note 7 and Supplementary Video 2. Compared with metasurface-based SCNN, pigment-based SCNN achieved mass production by lithography, thus obtaining high integration and high spatial resolution. However, the metasurfaces can provide more powerful light field modulation capabilities and greater design freedom[37-40], resulting in more spectral information and more space for customization. Based on the concept of the SCNN, the metasurface-based architecture also has further potential in sensing and processing other light dimensions, e.g., polarization and phase[43-45]. Besides, metasurfaces also have the potential to achieve mass production via standard semiconductor lithography process. Therefore, in practical, we can choose and design the optimal SCNN chip depending on the specific requirements of the application. It can be predicted that SCNN chips will have more potential in various applications.

## Discussion

We proposed an integrated SCNN framework that achieves in-sensor edge computing of incoherent natural light. It can detect visual information in natural raw 3D datacube with both spatial and spectral features by performing optical analog computing in real time. Leveraging both the OCL and ENLs, SCNN can achieve high performance even on edge devices with limited computing capabilities, which enables edge computing with MMI functions. In practical applications, utilizing the high versatility of ENLs, a specific SCNN chip can be easily adapted to various advanced vision tasks as demonstrated in this work. For the OCL, it is designed to perform inferencing for spectral sensing and computing in edge devices rather than in-situ training. Therefore, for a specific application, the weights can be fixed. It is a tailored chip for a specific task for edge computing applications. The computing speed and power consumption of OCL depend only on the exposure time and the power of the CIS, empowering ultrafast optical computing at high energy efficiency.

To achieve hyperspectral imaging and sensing, we can also adopt a conventional hyperspectral camera to scan hyperspectral images, and then process the images on GPU. However, such a system cannot be integrated on edge devices because GPU has large size, high energy consumption, and high cost that cannot meet the requirements of edge devices with limited computing capabilities. Besides, the conventional hyperspectral camera is also bulky, expensive, and not capable of real-time imaging. Our OCL is in-sensor computing that provides a computing speed as high as 21.0 TOPS and a substantial reduction of 96% in data throughput (Supplementary Note 2). Therefore, the SCNN makes it possible to process hyperspectral images using only a few extra digital neural network layers on edge devices. It can empower edge devices with both sensing and computing capabilities for various real-world complex vision tasks.

Compared with existing on-chip works, as is shown in Fig. 1, our SCNN can process natural hyperspectral images with high spatial resolution/pixels. It does not rely on coherent light source, fiber coupling, or waveguide delay. Although CIS is relatively slow (sampling rate is about 37 kHz) compared with the commonly used high-speed photodetector (sampling rate can exceed 100 GHz), we still achieve considerable computing speed and density compared with existing photodetector-based works because CIS has high integration and can take full advantages of space division multiplexing. If we replace the CIS with PD array, the computing speed can be further improved to over $10^7$ TOPS. Actually, as CIS is the most integrated optoelectronic device, we can have hundreds of millions of pixels at a very low cost. The SCNN provides the strategy of utilizing every single pixel to perform optical computing via CIS to achieve high computing density and reduce the number of photoelectronic conversions. Based on the above advantages of SCNN architecture, we have achieved mass production on a 12-inch wafer of the pigment-based SCNN, which still has the computing speed of over $10^{13}$ OPS (see Supplementary Note 2 for details). Thus, the proposed SCNN opens a new practical in-sensor computing platform for complex vision tasks with MMI functions in the real world.

**Table 1 Comparison with existing on-chip ONN works**

| Publication | Pixels | Computing speed | Computing density | In-sensor | Incoherent light | MMI | Application |
|---|---|---|---|---|---|---|---|
| X., X. et al[27] Nature, 2021 | 500×500 | 1.785 TOPS | - | × | × | × | handwritten digits recognition (HDR)/image processing |
| F., J. et al[26] Nature, 2021 | 128×128 | 4 TOPS | 1.2 TOPS/mm2 | × | × | × | HDR/edge detection |
| A., F. et al[11] Nature, 2022 | 5×6 | 0.27 TOPS | 3.5 TOPS/mm2 | × | × | × | low-resolution image classification |
| F., T. et al[24] Nat. Commun., 2023 | 28×28 | 13.8 POPS | - | × | × | × | HDR |
| M., X. et al[31] Nat. Commun., 2023 | 28×28 | 0.27 TOPS | 25.48 TOPS/mm2 | × | × | × | HDR |
| B., B. et al[32] Nat. Commun., 2023 | 250×250 | - | 1.04 TOPS/mm2 | × | × | × | HDR/edge detection |
| D., B. et al[33] Nature, 2024 | 28×28 | 0.108 TOPS | - | × | × | × | HDR |
| Ours | 400×533 | 21.0 TOPS | 5.3 TOPS/mm2 | √ | √ | √ | complex tasks in the real world: face anti-spoofing and disease diagnosis |

## Funding

This work is supported by the National Key Research and Development Program of China (2023YFB2806703, 2022YFF1501600). The National Natural Science Foundation of China (Grant No. U22A6004); Beijing Frontier Science Center for Quantum Information; and Beijing Academy of Quantum Information Sciences.

## Acknowledgment

The authors would like to thank Prof. Chengwei Qiu for his valuable discussion on the manuscript. And thank Dr. Jiawei Yang, Dr. Jian Xiong and Chenxuan Wang for their help and advice on this study.

## Author contributions

K. C., S. R. contributed equally to this work. K. C. conceived the study and proposed the optical computing strategy. S. R. completed the SCNN framework and conducted the experiments with the help of S. X.. Y. H. supervised the project and advised on device optimization. X. C., Z. H., and Y. W. completed the chip fabrication. S. W. and Y. L. advised on network design and training. S. R. and K. C. wrote the manuscript with contributions from all other coauthors. F. L., X. F., and W. Z. provided useful commentary on the results. All authors read and approved the manuscript.

## Methods

**Fabrication of the metasurface-based SCNN Chip.** The designed metasurfaces were formed using EBL on a silicon-on-insulator (SOI) chip. The silicon layer was 220 nm thick. The metasurface patterns were transferred onto the silicon layer via inductively coupled plasma etching (ICP). To remove the silicon layer from the underlayer, buffered hydrofluoric acid was used to wet etch the silicon dioxide layer. Finally, the entire top Si layer with the designed metasurfaces was transferred and attached to the surface of the CIS using polydimethylsiloxane (PDMS). We used a Thorlabs CS235MU camera for CIS. The proposed SCNN chip can be fabricated using a CMOS-compatible process and can be mass produced at low cost.

**Fabrication of the pigment-based SCNN Chip**. The SCNN Chip is produced at semiconductor foundry on a 12-inch wafer, employing a standard color filter array process via I-line lithography. The CIS wafer is uniformly coated with a color resist. To render the pattern insoluble, it is UV cured by exposure through a carefully designed photomask. Subsequent to this, any unnecessary portions of the color resist are removed using the developing solution. Following this removal, the pattern is further solidified through a baking process. This comprehensive sequence of steps is repeated nine times.

Following the color filter layer process, a planarization layer was established using the Chemical Mechanical Polishing (CMP) technique to ensure a flat and uniform surface. Subsequently, a photoresist layer was uniformly applied onto this planarized surface using a spin-coating method. This photoresist layer was patterned by UV light exposure through a predefined mask. The excess photoresist was then removed in a development process, leaving behind the desired patterns. The wafer was subjected to a reflow baking process, during which the patterned photoresist naturally reflowed into the shape of micro lenses, driven by surface tension and thermodynamic effects.

**Implementation of the ENNs.** The ENLs in the SCNN are realized using the TensorFlow[46] framework and trained on an NVIDIA RTX3080 GPU. Additional implementation and training details of ENLs are provided in Supplementary Note 6 and 7. After training, the ENLs and OCL formed a fully functional SCNN. The electrical components of the SCNN were run on an Intel Core i7-11700 @2.5GHz CPU for real-time applications.

# Supplementary Document for Spectral Convolutional Neural Network Chip for In-sensor Edge Computing of Incoherent Natural Light

Kaiyu Cui[#*1], Shijie Rao[#1], Sheng Xu[1], Yidong Huang[*1], Xusheng Cai[2], Zhilei Huang[2], Yu Wang[2], Xue Feng[1], Fang Liu[1], Wei Zhang[1], Yali Li[1], and Shengjin Wang[1]

[#] These authors contributed equally to this work

[*] Correspondent author: kaiyucui@tsinghua.edu.cn,

yidonghuang@tsinghua.edu.cn

Affiliation: [1]Department of Electronic Engineering, Tsinghua University, Beijing, China

[2]Beijing Seetrum Technology Co., Beijing, China

**Supplementary Note 1. Comparison between SCNN and on-chip snapshot hyperspectral imaging.**

On-ship snapshot spectral imaging (SSI) strategy needs to design a universal SSI chip for arbitrary spectrum reconstruction. It is based on compressive sensing theory. Therefore, SSI usually needs dozens to hundreds of different optical filters such as metasurface or Fabry-Pérot structures to get more compressive measurements and thus obtain higher spectrum reconstruction precision. Its applications focus on measuring spectra.

Take face anti-spoofing (FAS) as an example. SCNN is much more effective and practical. The comparison between our previous SSI-based method[1,2] and SCNN are listed below.

**Supplementary Table 1: Comparison between SSI-based and SCNN-based FAS**

|  | SSI[1,2] | SCNN |
|---|---|---|
| **Number of different metasurfaces** | 49~400 | 9 |
| **Spatial resolution (pixels)** | $5 \times 10^4$ | $\mathbf{2.13 \times 10^5}$ |
| **Frames per second (fps)** | <0.1 | >10 |
| **FAS accuracy (%)** | ~95% | ~99% |
| **Spectrum reconstruction** | √ | × |
| **Pixel-level real-time sensing** | × | √ |

SCNN aims at designing an application-oriented chip for real-world computer vision tasks based on optical neural network (ONN). It provides an in-sensor computing and non-reconstruction spectral imaging method for the final target of the downstream task. SCNN can use minimal metasurface units by just extracting the spectral features for specific applications. This provides an ONN-based approach for hyperspectral sensing tasks, effectively avoiding the need for as many metasurface units as possible for high precision spectral reconstruction. Further, fewer kernels enable higher feature compression capability, higher spatial resolution, and extremely lower computing costs for ENLs for the ONN with an optoelectronic framework. Generally, SSI systems need massive computing resources to complete the reconstruction procedure. Powered by modern artificial neural network technics and high-performance computing devices such as GPU, SSI systems are barely to achieve video-rate[3,4]. However, SCNN can reach video-rate easily even on a common laptop CPU, which empowers edge computing for terminal devices with limited computing capabilities.

**Supplementary Note 2. Computing speed of optical convolutional layer.**

The speed of analog computing cannot be simply quantified by operations per second (OPS). However, for a comparison with digital computing, we provide a method for calculating the equivalent OPS of our analog OCL based on the properties of our device. Notably, the computing speed is mainly limited by the pixels and the exposure time of the CMOS image sensor (CIS).

Considering a convolutional kernel of shape $n \times n \times C$, at each spatial location, the kernel will perform $n^2$ dot-product operations and $n^2$ summing operations. Each dot-product operation requires $C$ multiplications and $C-1$ summing operations. In all, we need $n^2(2C-1) + n^2 = 2n^2C$ operations. The final summing of the dot-product results only accounts for $\frac{n^2}{2n^2C} = \frac{1}{2C}$ of the computing burden. As $C$ is usually a large number, the computing burden of the summing of the dot-product results can be neglected. Moreover, this summing operation can also be completed by binning during the readout process of the image sensor. Therefore, we only take operations performed by optical computing into account.

For each pixel combined with spectral filter, assuming that the number of spectral sampling points is $C$, exposure time is $T$, then it can perform $C$ multiply operations and $C-1$ additive operations, resulting in the computing speed of $(C+C-1)/T \approx 2C/T$. For an OCL with a spatial resolution of $N$ pixels, the computational speed of the entire OCL can be calculated as follows:

$$s = N \cdot \frac{2C}{T} = \frac{2NC}{T}$$

In previous works, hyperspectral sensing with 601 sampling points (from 450~750nm at 0.5nm intervals) is realized using only 25 spectral filters[1,5]. That is, the compression rate is about 4.16%. Under this compression rate, using 9 spectral filters, the sampling points in the spectral dimension is about 216 ($C = 216$). In our implementation, the optical convolutional layer (OCL) has nine convolutional kernels of size 1 and stride 1. For the metasurface-based SCNN, the 3D raw data cube of the natural images has $160 \times 122$ superpixels ($480 \times 366$ pixels). The CIS used in our experiment was a Thorlabs CS235MU equipped with a Sony IMX249 sensor. The minimum exposure time is $0.034\ ms$. Therefore, $N = 480 \times 366, C = 216, T = 0.034 ms$, then the computing speed of the OCL is about 2.2 TOPS.

For the pigment-based SCNN, it has much higher integration and more spatial pixels. The 3D raw data cube of the natural images has 400×533 superpixels. If we process the raw hyperspectral image of 400×533×216 on electrical computing platform, we need about 176MB storage (stored in 32-bit floating point). The minimum exposure time is $0.027 ms$. Therefore, $N = 1200 \times 1098, C = 216, T = 0.027 ms$, then the computing speed of the OCL is about 21.0 TOPS and the OCL can reduce the storage requirement to 7.3MB. The SCNN provides a simple but highly effective way to sense and process hyperspectral images for various portable terminals. Notably, the pixel size of pigment-based SCNN is only $1.75 \times 1.75 \mu m$, resulting in the computing density of about 5.3 TOPS/mm². Moreover, the exposure time of the CIS is relatively low (sampling rate is about 37 kHz) compared with high-speed photodetector (sampling rate can exceed 100 GHz). If we replace the CIS with PD array, the computing speed can be further improved to over $10^7$ TOPS.

## Supplementary Note 3. Gradient-based metasurface topology optimization (GMTO) algorithm.

We first adopted freeform-shaped meta-atom metasurfaces[5] to generate millions of different metasurface units and arranged all the metasurfaces into a 2D array. Thus, each metasurface unit can be uniquely represented by a pair of coordinates $(p, q)$. To design $N$ metasurfaces, the objective can be considered a function of $2N$ independent variables: $L(p_1, q_1, \ldots, p_N, q_N)$.

Each metasurface can be described by its period $p \in [350, 550]$ and shape index $q \in [1, 10000]$. Every index $q$ can be mapped to a unique shape $S(q) \in R^{128 \times 128}$. Considering $N$ metasurfaces $A_{p_k, q_k}, k = 1, 2, \ldots, N$, each $A_{p_k, q_k}$ has the transmission response $t_{p_k, q_k} \in R^{M \times 1}$. Then, the correlation loss $L_{corr}$ can be calculated as follows:

$$L_{corr} = \max_{i, j = 1, 2, \ldots, N} \{t_{p_i, q_i}^T t_{p_j, q_j}\}$$

In addition, we employed a commercial hyperspectral camera to capture 5000 spectra of the positive samples $X_l \in R^{5000 \times M}$ and 5000 spectra of the negative samples $X_s \in R^{5000 \times M}$. Specifically, positive and negative samples represent live and spoof faces in FAS tasks and normal and pathological tissues in the disease diagnosis tasks, respectively. For each $A_{p_k, q_k}$, we utilized the inter-class variance and intra-class variance to quantify its anti-spoofing ability:

$$d_{p_k, q_k} = \sigma(X_l t_{p_k, q_k}) + \sigma(X_s t_{p_k, q_k}) - [mean(X_l t_{p_k, q_k}) - mean(X_s t_{p_k, q_k})]^2$$

where $\sigma$ denotes the variance and $mean$ denotes the mean value. Then, the FAS loss can be expressed as follows:

$$L_{fas} = \overline{d_{p,q}} = \frac{1}{N} \sum_{k=1}^{N} d_{p_k, q_k}$$

Because the fabrication precision of metasurface nanostructures is limited, we prefer to avoid fabricating two metasurfaces with similar shapes or periods. Although the simulated transmission responses of these two metasurfaces may have a low correlation, the fabrication error may result in a high similarity between the two actually fabricated metasurfaces. Therefore, we also introduce the fabrication loss into the final loss function to avoid designing similar metasurfaces. The similarity between the two metasurfaces is calculated as follows:

$$sim\left(A_{p_i,q_i}, A_{p_j,q_j}\right) = s(q_i)^T s(q_j) - \log\left(1 + 0.005 \times abs(p_i - p_j)\right)$$

where $s(q) \in R^{16384 \times 1}$ represents the flattened value of $S(q) \in R^{128 \times 128}$, $\log$ represents a logarithm with a base of 10, and $abs$ represents the absolute value. The fabrication loss can be calculated as follows:

$$L_{fab} = \max_{i,j=1,2,\ldots,N}\{sim\left(A_{p_i,q_i}, A_{p_j,q_j}\right)\}$$

Finally, the total loss was calculated and minimized to obtain an optimized metasurface design.

$$L_{total} = \alpha L_{fas} + \beta L_{corr} + \gamma L_{fab}$$

$$\{p_k, q_k\} = \arg\min_{\{p_k, q_k\}} L_{total}$$

**Supplementary Note 4. Test results for FAS.**

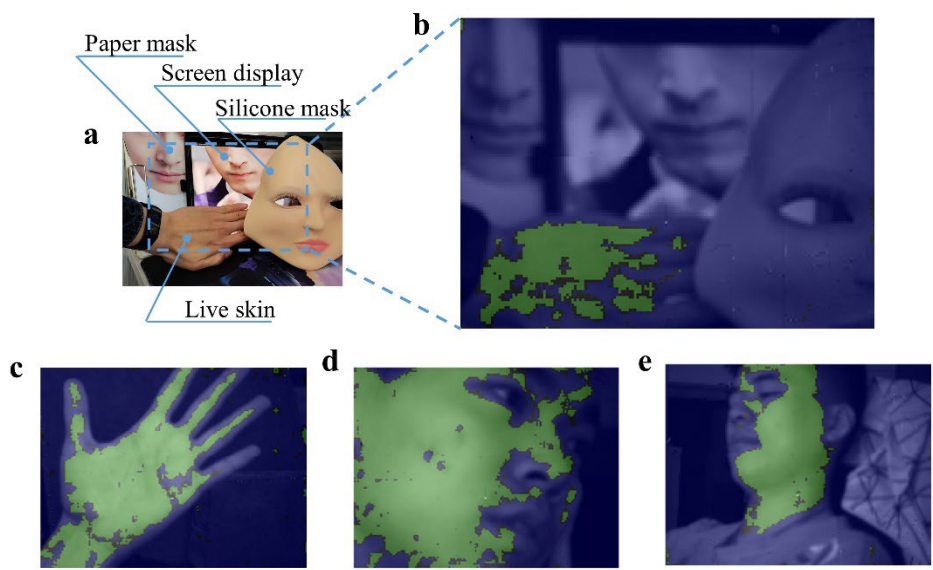

**Supplementary Fig. 1. Real world test results for pixel-level anti-spoofing liveness detection. a,** A test scene consists of a live human hand and several presentation attacks, including paper mask, screen display, and silicone mask. **b,** Predicted results of the test scene by spectral convolutional neural network (SCNN), where the green points represent the live human skin area. **c-d,** Predicted results of other real world scenes.

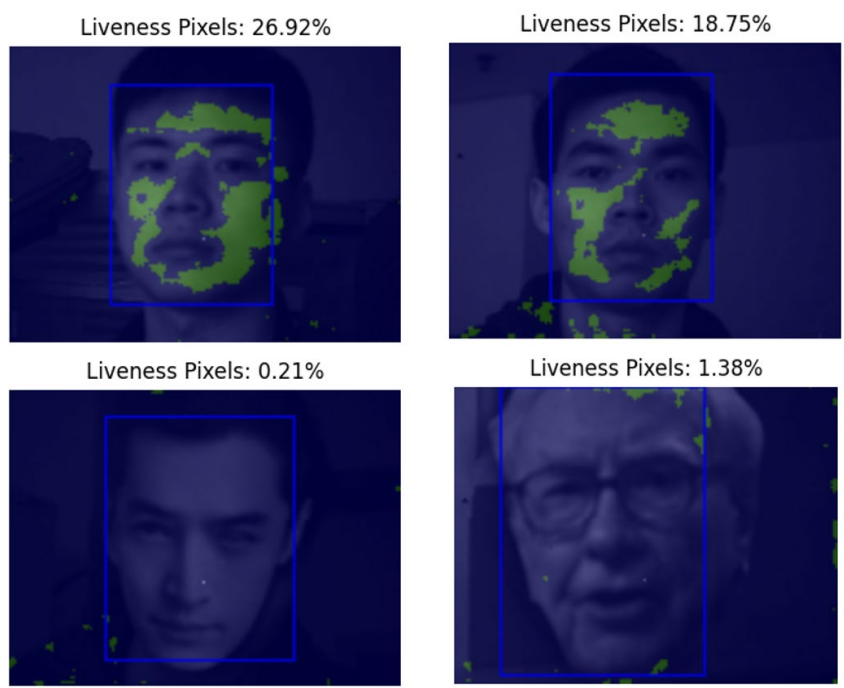

**Supplementary Fig. 2. SCNN for image-level FAS on four testing samples.**

By calculating the liveness area ratio (the proportion of liveness pixels in the detected face bounding box), SCNN can perform reliable image-level FAS. In our experiment results, the ratio of live face image is generally greater than 15% and the ratio of spoof face image is generally less than 1.5%. Therefore, by further processing the pixel-level liveness detection results, this difference in order of magnitude enables the 100% accuracy of FAS. Moreover, by changing and retraining the ENLs, the SCNN can directly achieve 100% accuracy of image-level anti-spoofing.

**Supplementary Note 5. Design of ENLs for anti-spoofing face recognition.**

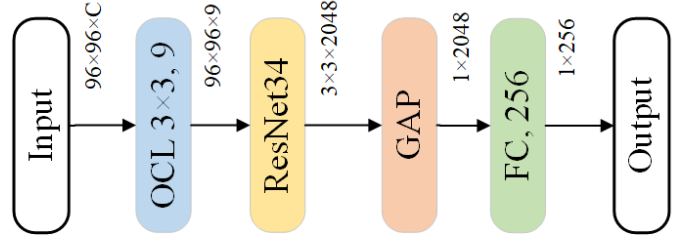

**Supplementary Fig. 3. Network architecture of the proposed SCNN for face recognition.** The OCL layer sensors and simultaneously calculates the convolution results of the hyperspectral facial image. Then ResNet34[6] is adopted to further extract spatial and spectral features from the outputs of OCL layer. Finally, the hyperspectral facial image is embedded into a latent vector $f \in R^{1\times 256}$. We pretrained the network using Arcface[7] loss and stochastic gradient descent optimizer on MS1M[8] dataset. Then we fine-tuned the SCNN on the dataset captured by our own sensor.

Here, we demonstrate the way to use the SCNN to perform complex vision tasks other than image classification, considering face detection and recognition as examples. In the face recognition task, the OCL inputs were $96 \times 96 \times C$ data cubes. Here $C$ denotes the number of spectral channels and the ENL inputs were $96 \times 96 \times 9$ data cubes. In the face-detection task, the OCL and ENL input sizes were $96 \times 128 \times C$ and $96 \times 128 \times 9$, respectively.

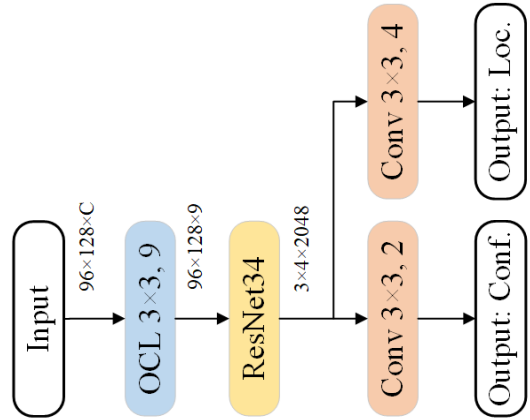

**Supplementary Fig. 4. Network architecture of the proposed SCNN for hyperspectral face detection.** The function of OCL layer and ResNet34 are the same as described in Supplementary Fig. 3. However, on the outputted feature maps of ResNet34, 2 convolution kernels of size $3 \times 3$ are used to calculate the confidence

(Conf.) of predicted label. Another four convolution kernels of size $3 \times 3$ are used to predict the location (Loc.) of bounding boxes. SSD[9] loss was adopted to train the network. The network was first trained on Wider Face[10] dataset and then fine-tuned on the dataset captured by our sensor. These works show that by simply changing the network layers implemented on CPU/GPU, SCNN can perform various advance computer vision tasks on hyperspectral image.

**Supplementary Note 6. Design and training details of the ENLs for metafurface-based SCNN.**

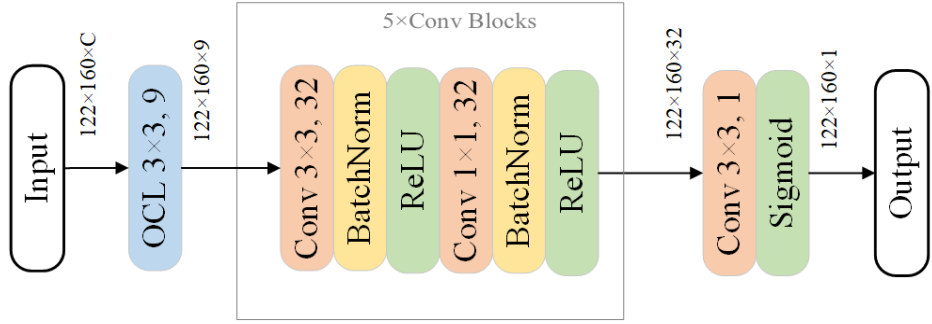

**Supplementary Fig. 5. Network architecture of our SCNN for pixel-level disease detection and liveness detection.**

The inputs of SCNN are the 3D raw data cube of natural images. The OCL layer sensors and calculates the results of convolution at the same time. It has nine convolutional kernels of size three and stride nine. The feature maps outputted by OCL are further processed by the following electrical layers on CPU. The final outputs of $122 \times 160 \times 1$ represents the pixel-level detection results.

To train the ENLs for disease detection, we employed our SCNN chip to capture 2990 samples of thyroid histological sections through a microscope and randomly selected 250 samples to serve as the test set, using the rest as the training set. To train the ENLs for FAS, we employed our SCNN chip to capture more than 200 samples of live skin and spoof material in the real world. We then obtained pixel-level annotations of the feature cubes by manual labeling. Pixels located on live human skin were labeled positive, whereas non-live pixels on various materials, including environmental objects and spoof materials such as silicone masks, latex masks, and resin masks, were labeled negative. Finally, ENLs were trained on the labeled dataset. Then we employed our SCNN chip to capture images, obtained a test set containing 108 test samples from 31 individuals, and evaluated the performance of the SCNN framework.

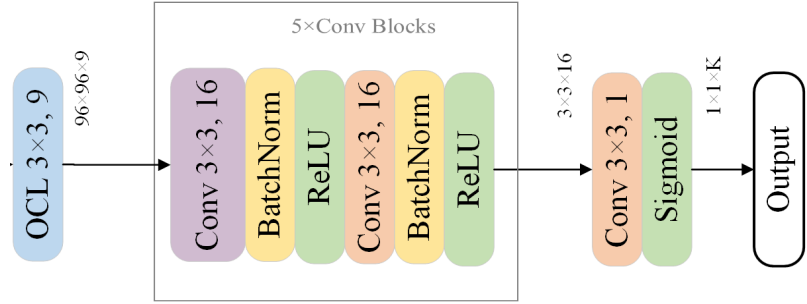

**Supplementary Fig. 6. Network architecture of our SCNN for image-level disease diagnosis and FAS.**

The inputs of SCNN are 3D raw data cube of natural images with size $96 \times 96 \times C$, where $C$ denotes the number of spectral channels. The OCL has nine convolutional kernels of size three and stride nine. Therefore, the feature maps outputted by OCL has the size $96 \times 96 \times 9$. In the ENLs, each convolutional block contains two convolutional layers with 16 kernels of size three, two BatchNorm layers, and two ReLU layers. The strides of the two convolutional layers are one and two. The final outputs of $1 \times 1 \times K$ represents the pixel-level detection results. Here, $K$ represents the number of classes, which is five in thyroid disease diagnosis task and two in FAS task.

**Supplementary Note 7. Experimental results of pigment-based SCNN chip.**

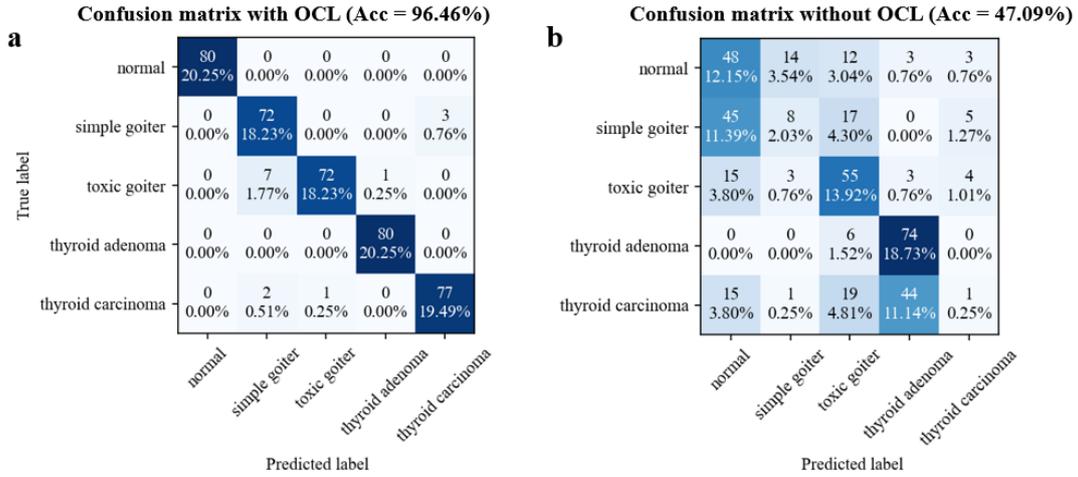

**Supplementary Fig. 7 Experimental results of thyroid histological section diagnosis by the SCNN. a,** Our SCNN achieves an accuracy of 96.46% on the thyroid disease diagnosis task. **b,** The classification accuracy is only 47.09% without the OCL.

For FAS, we collected more than 200 samples of live subjects (from 5 different people) and spoof subjects (from 15 different masks). Then the dataset was split into training set and testing set at a ratio of 4:1 according to the subject identities. The confusion matrix of the classification results on the testing dataset is shown in Supplementary Fig. 7a. Note that all of the misclassified samples are between the four diseases, which indicates that the four thyroid diseases are not completely independent and that there may be complicating pathologies. Moreover, if we distinguish only normal samples from diseased samples, the accuracy on the testing dataset is 100%. Furthermore, we conducted another experiment by replacing OCL with CIS without pigment-based filters to study the role of OCL. After repeating the same data collection and ENL training procedure, the classification accuracy decreased from 96.46% to 47.09%. This indicates that spectral information is vital to diagnosis. Our OCL enables powerful spectral sensing capabilities, and the features acquired by OCL are effective for the precise diagnosis of pathological sections. Moreover, the diagnosis process does not require a microscope.

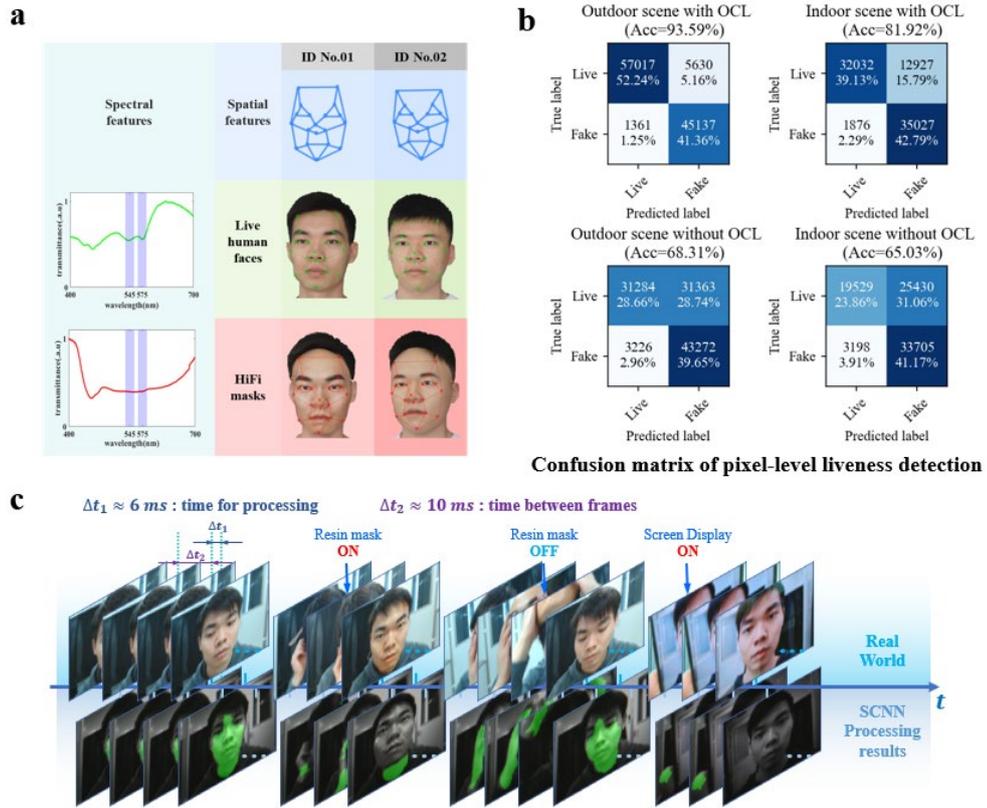

**Supplementary Fig. 8 SCNN chip can be used for pixel-level anti-spoofing liveness detection. a,** Our SCNN chip can combine spectral features with spatial features and perform reliable anti-spoofing face recognition. **b,** Confusion matrix for the pixel-level liveness detection results with and without OCL under sunlight and white LED light. **c,** Video-rate liveness detection based on the SCNN chip can detect high-fidelity lifelike masks effectively. The detected pixels of live skin are marked in green. More results can be found in the Supplementary Video.

In addition to histological section diagnosis, we employed the proposed SCNN for FAS to further study its capability for computer vision tasks. Nearly all of the current face recognition systems can be deceived by high-fidelity (HiFi) silicone masks, posing a great risk to privacy and security. However, discriminative features can be extracted to detect HiFi masks when powered by our MMI. The broadband incoherent natural light that includes two spatial dimensions and one spectral dimension that are first captured and processed by the OCL. The size of the feature maps output by OCL is 400 × 533 × 9. Then, the feature maps are further processed by several ENLs, and we can obtain the anti-spoofing pixel-level liveness detection results. Our SCNN chip can combine spectral features with spatial features and perform reliable anti-spoofing face recognition (Supplementary Fig. 8a). To test its real-world performance, we collect data samples in both outdoor scene and indoor scene. Supplementary Fig. 8b shows the confusion matrix for the test samples. The SCNN achieves accuracies of 93.59% and 81.92% in outdoor and indoor scenes, respectively. In the outdoor scene, the ambient light source is sunlight. Because sunlight covers a wider spectral band, it also has a high intensity in the near-infrared band, which is useful for anti-spoofing and leads to better results. The indoor scene is mainly illuminated by artificial light sources such as LED. They usually have weak intensity in the near-infrared band. To improve the performance in the indoor scene, we can adopt a wide-band lamp as the fill light. And we will also further optimize the hardware design specifically for indoor scenes. If we remove the OCL, the accuracies decrease to 68.31% and 65.03%, respectively, which demonstrates that the spectral information obtained by OCL is of vital importance.

Furthermore, we can conduct image-level FAS based on the pixel-level liveness detection results. By applying an additional face detection procedure, we can get the bounding boxes of the faces and then calculate the averaged value of the pixel-level liveness detection results for each bounding box. As for the live and spoof faces shown in Fig. 4g, the averaged values are 0.6052 and 0.0016 respectively. In this way, we can get almost 100% image-level anti-spoofing accuracy. To show the real-world FAS capabilities, we employed the designed SCNN chip to perform real-time anti-spoofing pixel-level liveness detection at different video frames (Supplementary Fig. 8c). The frame rate of the results is almost only limited by the CIS exposure time. The HiFi masks can be easily detected at the pixel level (more results can be found in the Supplementary Video). Thus, the proposed SCNN framework is expected to be widely used in real-world MMI applications. The results indicate that by simply redesigning and retraining the ENLs according to the needs of specific tasks, the function of the SCNN can be customized as the disease diagnosis task and the liveness detection task performing at image and pixel levels. The final output of the SCNN is highly customizable. The SCNN can flexibly adapt to various advanced CV tasks at video rates by simply changing and retraining the ENLs. It can combine the advantages of optical and electrical computing.

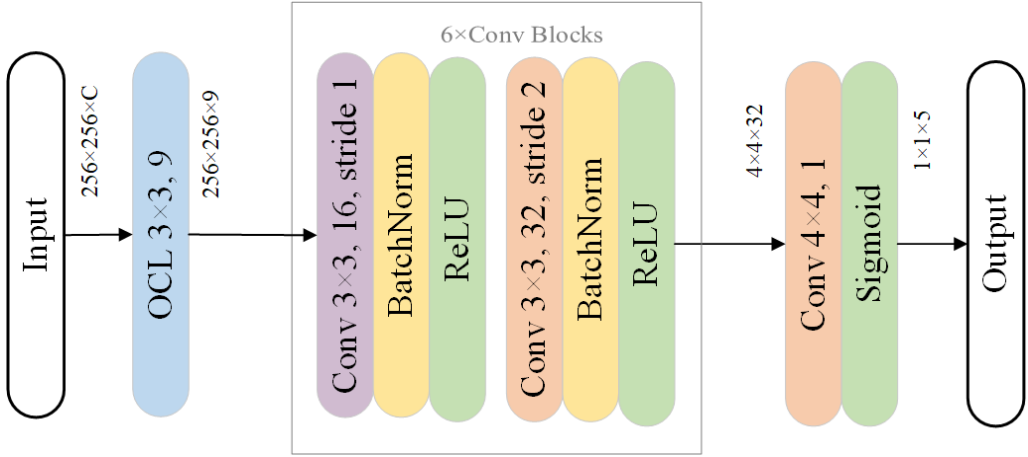

**Supplementary Fig. 9. Network architecture of our SCNN for image-level disease diagnosis.**

The inputs of SCNN are 3D raw data cube of natural images with size $256 \times 256 \times C$, where $C$ denotes the number of spectral channels. The OCL has 9 convolutional kernels of size $1 \times 1$ and stride 1. Therefore, the feature maps outputted by OCL has the size $256 \times 256 \times 9$. In the electrical network layers (ENLs), each convolutional block contains two convolutional layers. Each convolutional layer is followed by a BatchNorm layer and a ReLU layers. The first convolutional layer has 16 kernels of size $3 \times 3$ and stride 1. The second convolutional layer has 32 kernels of size $3 \times 3$ and stride 2 to perform downsampling. The final outputs of $1 \times 1 \times 5$ represents the 5-class classification results.

We utilize 100 histological sections from 100 different patients to collect the dataset. The 100 sections contain 5 categories (normal, simple goiter, toxic goiter, thyroid adenoma, and thyroid carcinoma), each with 20 sections. We collected about 500 samples from these 100 different sections. 80 sections were employed as training set and the remaining 20 sections were testing set. Then we employ our SCNN sensor to capture the 9-channel feature maps of these sections and build the training and testing set. Each section has been sampled several times. The raw outputs of the SCNN sensor have the size of $400 \times 533 \times 9$. We randomly crop the raw outputs to the size of $256 \times 256 \times 9$ to perform data augmentation. Finally, the ENLs are trained using the collected dataset. The Adam optimizer and cross-entropy loss are adopted to train the network and the learning rate is 0.001.

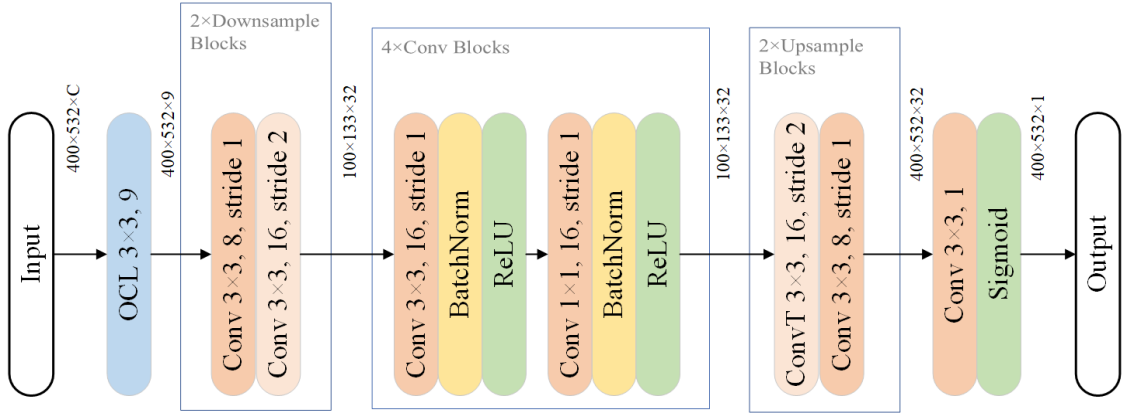

**Supplementary Fig. 10. Network architecture of our SCNN for pixel-level liveness detection.**

The feature maps outputted by OCL are further processed by the following electrical layers on CPU/GPU. After several convolutional layers, the final outputs have size $400 \times 532 \times 1$ and represent the pixel-level anti-spoofing liveness detection results. To train the ENLs for FAS, we employ our SCNN sensor to capture more than 200 samples of live skin and spoof material in the real world. The spoofing materials include silicone masks, paper masks, resin masks, and screen display. We then obtain pixel-level annotations of the feature cubes by manual labeling. Pixels located on live human skin are labeled as positives, whereas non-live pixels on various materials, including environmental objects and spoof materials are labeled as negatives. Finally, ENLs are trained on the labeled dataset. The ENLs can achieve a processing speed of about 14 frames per second (fps) at a batch size of 1, or about 20 fps at a batch size of 8, on an Intel Core i7-11700 @2.5GHz CPU. As the computing of OCL and ENLs can be asynchronous. That is, while the ENLs are processing the current frame, the OCL is capturing the next frame. In this way, the real-world performance of SCNN can achieve video rate.

**Supplementary Note 8. Additional analysis on disease diagnosis application of the pigment-based SCNN.**

The disease diagnosis is a 5-class classification task. Assuming that the confusion matrix $M = \{m_{ij}\} \in R^{5\times 5}$, each row of $M$ represents a true label and each column of $M$ represents a predicted label. $m_{ij}$ represents the number of testing samples that have true label $i$ and are predicted to be $j$. Therefore, the diagonal elements indicate the number of samples that were correctly categorized. The Acc (accuracy) metric is calculated as $\frac{tr(M)}{sum(M)}$, which indicates the overall classification accuracy. For the classification task, the other commonly used evaluation metrics besides accuracy are precision (calculated as $\frac{m_{kk}}{\sum_{i=1}^{5} m_{ik}}$) and recall (calculated as $\frac{m_{kk}}{\sum_{j=1}^{5} m_{kj}}$) for each class $k$. Here we provide the precision and recall for each class in the table below.

|  | Normal | Simple goiter | Toxic goiter | Thyroid adenoma | Thyroid carcinoma |
|---|---|---|---|---|---|
| **Precision** | 100% | 88.89% | 98.63% | 98.77% | 96.25% |
| **Recall** | 100% | 96.00% | 90.00% | 100% | 96.25% |